\documentclass[12pt,preprint,preprintnumbers,nofootinbib,amsmath,amssymb]{revtex4}

\usepackage[dvips]{color}
\usepackage{graphicx}
\usepackage{epsfig}    
\usepackage{dcolumn}
\usepackage{bm}
\usepackage{amssymb}
\usepackage{epsfig}    
\usepackage{here}

\newcommand{\lsim}{\mbox{ \raisebox{-1.0ex}{$\stackrel{\textstyle <}
{\textstyle \sim}$ }}}

\def\Journal#1#2#3#4{{#1} {\bf #2} (#4) #3}

\def\PLB{{\em Phys. Lett.}  B}
\def\PRL{\em Phys. Rev. Lett.}
\def\PREP{\em Phys. Rep.} 
\def\PRD{{\em Phys. Rev.} D}

\def\MPLA{{\em Mod.~Phys.~Lett.} A}

\begin{document}
\topmargin -1cm

\begin{flushright}
\today \\
\end{flushright}

\begin{center}
{\large \bf 
Search for Lepton Flavor Violation 
in the Higgs Boson Decay\\ at a Linear Collider}

\vspace*{10mm}
{\sc Shinya Kanemura}\footnote{
E-mail: kanemu@het.phys.sci.osaka-u.ac.jp}, 
{\sc Kouichi Matsuda}\footnote{
E-mail: matsuda@het.phys.sci.osaka-u.ac.jp},  
{\sc Toshihiko Ota}\footnote{
E-mail: toshi@het.phys.sci.osaka-u.ac.jp} \\
{\sc Tetsuo Shindou}\footnote{Address after April 2004:
Theory Group, KEK, Tsukuba, Ibaraki 305-0801, Japan, \vspace{-0.1cm}\\
E-mail: shindou@post.kek.jp}, 
{\sc Eiichi Takasugi}\footnote{
E-mail: takasugi@het.phys.sci.osaka-u.ac.jp},  
{\sc Koji Tsumura}\footnote{
E-mail: ko2@het.phys.sci.osaka-u.ac.jp} \\   
\vspace*{4mm}
{\em Department of Physics, Osaka University, Toyonaka, Osaka
 560-0043, Japan}\\
\end{center}

\begin{abstract}
We discuss possibility of direct search for lepton flavor violation 
(LFV) in Yukawa interaction by measuring the branching ratio 
for the decay of the lightest Higgs boson ($h^0$) 
into a $\tau$-$\mu$ pair at a linear collider. 
We study the significance of the signal process,  
$e^+e^- \to Z^\ast \to Z h^0 \to Z \tau^\pm \mu^\mp$, 
against the backgrounds such as 
$e^{+}e^{-} \rightarrow Z \tau^{+}\tau^{-} \rightarrow Z \tau^{\pm}\mu^{\mp}+$
missings.
After taking appropriate kinematic cuts, the number of the background 
event is considerably reduced, so that the signal can be visible when 
the branching ratio of $h^0 \to \tau^\pm \mu^\mp$ 
is larger than about $10^{-4}$. 
In a Minimal Supersymmetric Standard Model scenario, 
the effective coupling of $h^0 \tau^\pm \mu^\mp$ can be 
generated at loop level due to the slepton mixing.
When 
supersymmetric mass parameters are larger than TeV scales,
the branching ratio can be as large as several times $10^{-4}$. 
Therefore, the signal can be marginally visible at a LC. 
In the general two Higgs doublet model, the possible maximal value 
for the branching ratio of $h^0 \to \tau^\pm \mu^\mp$ 
can reach to a few times $10^{-3}$ within the available 
experimental bound, so that we can obtain larger significance. 
\end{abstract}

\maketitle

\section{Introduction}

Lepton Flavor Violation (LFV) is a direct indication of 
new physics beyond the Standard Model (SM). 
It can naturally appear in a scenario based on the supersymmetry 
(SUSY) due to the slepton mixing.  
Its origin may be the radiative effect 
of the neutrino Yukawa interaction with heavy right-handed 
neutrinos \cite{rge1,rge2}. 
There are some other scenarios which naturally induce LFV, 
such as the Two-Higgs-Doublet Model (THDM) with general 
Yukawa interaction, the Zee model for neutrino masses\cite{zee} and so on.
In the low energy effective theory of such new physics models, 
two kinds of the Lepton Flavor (LF) violating couplings exist; i.e., 
those associated with gauge bosons and those with Higgs bosons 
(LF violating Yukawa couplings). 

In recent years, the Higgs-mediated LF violating 
processes have been studied regarding the decay modes   
$\tau^\pm \to \mu^\pm\mu^\pm\mu^\mp$ \cite{babu,ellis,rossiNew},
$\tau^{\pm} \rightarrow \mu^{\pm} \eta$ \cite{Sher},
and $B_s \to \mu^\pm\tau^\mp$ \cite{ellis}.    
Their branching fractions are being measured 
at current and forthcoming experiments at the B-factories
\cite{B-factories} and CERN Large Hadron Collider (LHC) \cite{LHC}.
The Higgs-mediated LFV have also been investigated in muon processes: 
the muon-electron conversion in nucleus is studied in Ref.~\cite{mu-e}.
It will be explored at the Muon to Electron COnversion experiment (MECO)
\cite{MECO}
and the PRISM Muon Electron conversion experiment (PRIME) based on
the Phase Rotate Intense Slow Muon (PRISM) \cite{PRISM}.
All these processes are measured as a combination of contributions
from the gauge boson mediation and the Higgs boson mediation.

In this letter, we consider possibility of detecting the process 
of the lightest Higgs boson decaying into a pair of tau and muon, 
$h^{0} \rightarrow \tau^{\pm} \mu^{\mp}$, at a Linear Collider (LC).
There LFV in Yukawa interaction can be directly studied 
by measuring the decay branching ratio of the Higgs bosons 
\cite{rossiNew, rossi, assamagan, HiggsLFVDecay} when 
they are found.
In the Minimal Supersymmetric Standard Model (MSSM), the mass of the 
lightest Higgs boson is less than about $130$ GeV.
It is promising that such a light Higgs boson will be discovered at LHC. 
Then its properties such as the mass, the width, production cross sections, 
and decay branching ratios will be measured extensively. 
The precision study of the Higgs sector is one of the main purposes 
of a LC such as GLC, TESLA or NLC\cite{lc}.
The lightest Higgs boson is produced mainly through gauge interactions
at a LC. In the case of the {\it nearly} decoupling region\cite{haber},  
the production cross section for the lightest Higgs boson
is much lager than that of the heavier ones. 
Therefore, the LF violating Yukawa coupling 
can be better tested from the decay of the the lightest Higgs boson 
than that of the extra (heavier) Higgs bosons.  

At LHC, the extra Higgs bosons ($H^{0}$, $A^{0}$, $H^\pm$) 
may also be detected in the MSSM and the THDM as long as 
their masses are not too large.
The Higgs bosons are mainly produced through the Yukawa interaction, 
so that the production cross section of $H^0$ and $A^0$ can be 
sufficiently large to be detected 
especially for large $\tan\beta$ values, 
where $\tan\beta$ is the ratio
of vacuum expectation values of two Higgs doublets. 
Therefore, the decays of $H^0$ and $A^0$ may be 
useful to explore the LF violating Yukawa coupling\cite{assamagan}.  
The search of LFV via the Higgs boson decay 
at a hadron collider is suffered from huge backgrounds, 
and it should be required to pay much effort into the 
background reduction.

Magnitude of the LF violating coupling 
in $h^{0} \rightarrow \tau^{\pm}\mu^{\mp}$
is constrained by the results from the measurement of LFV in 
$\tau$ decay processes.
The most stringent bound comes from the 
$\tau^{-} \to \mu^{-}\eta$ measurement\cite{expTauMuEtaBelle}.
In the framework of the MSSM, the theoretical prediction on the branching
ratio of $h^{0} \to \tau^\pm\mu^\mp$ can approach to the above experimental 
upper limit by adjusting the SUSY parameters\cite{rossi}; i.e. 
${\rm Br}(h^0 \to \tau^\pm \mu^\mp) \sim$ several times $10^{-4}$. 
When all the SUSY parameters are as large as TeV scales, 
the LF violating gauge-boson penguin diagram decouples 
from the experimental reach, while the LF violating Yukawa coupling does not
because they depend only on the ratio of the SUSY parameters.
We can then avoid strong correlation between the LFV mediated 
by Higgs bosons and that by the gauge bosons.
On the contrary, if the scale of the SUSY parameters is smaller 
than 1 TeV, the Higgs-mediated LF violating coupling is  
strongly constrained from the experimental bounds 
on the gauge mediated LFV processes \cite{ellis}. 
In such a case, the parameter choice which realizes 
${\rm Br}(h^0 \to \tau^\pm\mu^\mp) \sim {\cal O} (10^{-4})$ is already 
excluded by the data.

We evaluate the significance of detecting the signal for 
$h^{0} \to \tau^\pm \mu^\mp$ at a LC. 
The Higgs boson with the mass around 120 GeV 
is mainly produced through the Higgsstrahlung 
mechanism $e^{+} e^{-} \rightarrow Zh^{0}$, 
when the center-of-mass energy $\sqrt{s}$ is lower than about 500 GeV. 
We can identify the signal event ($\tau^\pm\mu^\mp Z$)
without measuring the $\tau$ lepton  
by using the information of the momenta for the outgoing 
$Z$ boson and muon as well as the fixed beam energy $\sqrt{s}$.
The momentum of the $Z$ boson is reconstructed from those 
of its leptonic ($\ell^+\ell^-$ with $\ell^\pm=e^\pm$ and $\mu^\pm$) 
as well as hadronic  ($jj$) products. 
The most serious  irreducible background is 
$e^{+}e^{-} \rightarrow Z h^{0} \to Z \tau^{+} \tau^{-}$ 
with one of the tau leptons going to a muon and missings.  
The background can be suppressed by appropriate kinematic cuts 
with the expected resolution of the momentum of the $Z$ boson  
from the decay channels into $\ell^+\ell^-$ and $jj$ 
and with the beam spread rate of $\sqrt{s}$. 
We find that the significance $S/\sqrt{B}$ can 
exceed 5 in the MSSM scenario when the SUSY parameters 
are taken to be as large as TeV scales.
In the general THDM, the larger number of the signal events 
can be realized under the constraint from 
the perturbative unitarity\cite{pu1,pu2}, the vacuum stability\cite{vs} 
and available data. 
Therefore, the signal can be marginally detectable 
in the MSSM.

In Sec.~II, the possible enhancement of 
the decay branching ratio for the process 
$h^0 \to \tau^\pm \mu^\mp$ is discussed taking into 
account the current experimental data.  
We show a choice of the SUSY parameters that realizes a relatively 
large value of the effective $h^0 \mu^\pm \tau^\mp$ coupling in the MSSM.
In Sec.~III, we estimate the significance of detection 
for the signal against the backgrounds at a LC, taking 
into account appropriate kinematic cuts.
The conclusions are given in Sec.~IV.

\section{Lepton Flavor Violating Yukawa Coupling}

The effective Lagrangian of Yukawa interaction for charged leptons
in the THDM (including the MSSM) is described as
\begin{eqnarray} 
   {\cal L}_{\text{eff}} = - \overline{\ell}_R^i Y_{\ell_i}^{} 
                        \left(
                        \delta_{ij} \Phi_{1} 
                        + 
                        \epsilon_{ij} \Phi_{2} \right) \ell_L^j 
                   + {\rm h.c.}, 
\label{eq:L-Yukawa}
\end{eqnarray} 
where $\ell_{L,R}^i$ ($i=1,2,3$) are charged leptons with chirality 
$L$ or $R$, 
$\Phi_{\alpha}$ ($\alpha=1,2$) 
are neutral components of the two Higgs doublets 
with the hypercharge $1/2$, and 
$Y_{\ell_i} ( = m_{\ell_i}^{}/\langle \Phi_1^{} \rangle)$ 
are the Yukawa coupling constants of $\ell_i^{}$, respectively. 
In the MSSM, $\Phi_{1}$ and $\Phi_{2}$ correspond to 
$H_{d}^0$ and $H_{u}^{0*}$, respectively\cite{HHG}.   
With a nonzero value of $\epsilon_{ij}$ $(i \neq j)$, 
the Yukawa interaction and the mass of charged leptons 
cannot be diagonalized simultaneously, so that
the LF violating Higgs couplings arise. 
The interaction corresponds to $\tau$-$\mu$ or $\tau$-$e$ mixing 
is expressed \cite{babu,ellis, mu-e,rossi} by 
\begin{eqnarray} 
 {\cal L}_{\tau \ell_i} = - \frac{ \kappa_{3i} m_\tau}{v \cos^2\beta}
                 (\overline{\tau_{R}} \ell_{L i}) 
 \left\{ \cos(\alpha-\beta) h^0 + \sin(\alpha-\beta) H^0 - {\rm i} A^0 
\right\} + \text{h.c.}, 
\label{lfvc}
\end{eqnarray} 
with $\ell_{L_i}=e_L^{}$ or $\mu_L^{}$, and the LF violating parameter 
$\kappa_{ij}$ is given by 
\begin{eqnarray} 
 \kappa_{ij} = - \frac{\epsilon_{ij}}
               {\left(
                 1 
                 +
                 \epsilon_{33} \tan\beta \right)^2}, 
\end{eqnarray} 
where $h^{0}$ and $H^{0}$ are the CP-even Higgs bosons,  
$A^{0}$ is the CP-odd Higgs boson,
$\alpha$ denotes the mixing angle of the CP-even Higgs bosons, and 
$\tan\beta$ is the ratio of the vacuum expectation
values, $\tan \beta \equiv \langle \Phi_{2} \rangle / \langle \Phi_{1}
\rangle$.
We define $h^{0}$ is lighter than $H^{0}$ ($m_{h} < m_{H}$). 

Let us discuss the branching ratio of the Higgs boson decaying  
into the LF violating channel ($\tau^\pm\mu^\mp$).
We consider the situation that the main decay mode of the lightest 
Higgs boson is $h^{0} \to b \overline{b}$. 
In addition, for a large $\tan\beta$ and $\sin(\alpha-\beta) \simeq -1$, 
the dominant decay modes of heavier Higgs bosons are 
those into a $b \overline b$ pair.
In this case, 
the rate between the decay widths of the lepton flavor violating 
process $\Phi^{0} \to \tau^{+}\mu^{-}$ $(\Phi^{0} = h^{0},H^{0},A^{0})$ and
$\Phi^{0} \to b \overline{b}$ 
approximately gives the order of the branching ratio for 
$\Phi^{0} \to \tau^{+}\mu^{-}$; i.e., 
\begin{eqnarray} 
 \text{Br}(h^{0} \to \tau^{\pm}\mu^{\mp}) &\sim& \frac{1}{N_c}  \frac{m_\tau^2}{m_b^2} 
        \frac{\cos^2(\alpha-\beta)}{\cos^2\beta\sin^2\alpha} 
        \times |\kappa_{32}|^2, \label{eq:Br-h-TauMu}\\
 \text{Br}(H^{0} \to \tau^{\pm}\mu^{\mp}) &\sim& \frac{1}{N_c}  \frac{m_\tau^2}{m_b^2} 
        \frac{\sin^2(\alpha-\beta)}{\cos^2\beta\cos^2\alpha} 
        \times |\kappa_{32}|^2, \hspace{4mm} (\tan\beta \gg 1),\\
 \text{Br}(A^{0} \to \tau^{\pm}\mu^{\mp}) &\sim& \frac{1}{N_c}  \frac{m_\tau^2}{m_b^2} 
        \frac{1}{\sin^2\beta\cos^2\beta} 
        \times |\kappa_{32}|^2, \hspace{4mm} (\tan\beta \gg 1).
\end{eqnarray} 
In our numerical evaluation,
we calculate these branching ratios 
including all the decay modes 
i.e., $\Phi^{0} \to b\bar{b}$, $\tau^{+}\tau^{-}$, $c\bar{c}$, $gg$, $WW^{(\ast)}$, $ZZ^{(*)}$.  

The LF violating parameter $|\kappa_{32}|$ is constrained from 
the available data for LFV in $\tau$ decay processes.
The most stringent bound is obtained from 
the  $\tau^{-}\rightarrow \mu^{-}\eta$ measurement\cite{expTauMuEtaBelle}. 
The other experiments such as 
$\tau^{\pm} \rightarrow \mu^{\pm} \mu^{\pm}\mu^{\mp}$\cite{tau3mu}, 
$\tau^{\pm} \rightarrow \mu^{\pm}\gamma$\cite{taumugamma},
and muon anomalous magnetic moment give weaker bounds.
The branching ratio of $\tau^{\pm}\rightarrow \mu^{\pm}\eta$ is
given \cite{babu,ellis,Sher} by
\begin{align} 
  \text{Br}(\tau^\pm \xrightarrow{A^{0}} \mu^\pm \eta)
 &=
 8.4 \times \text{Br}(\tau^{\pm} \xrightarrow{\Phi^{0}}
 \mu^{\pm}\mu^{\pm}\mu^{\mp}) \nonumber \\
 &= 8.4 \times \frac{G_F^2 m_\mu^2 m_\tau^7 \tau_\tau}{1536 \pi^3 }
 \left( \frac{1}{m_{H}^{4}} + \frac{1}{m_{A}^{4}} \right)
     |\kappa_{32}|^2 \tan^6\beta,  
\end{align} 
for $\tan\beta \gg 1$ and $\sin(\alpha-\beta) \simeq -1$.
The present experimental bound is given by  
$\text{Br}(\tau^{-} \to \mu^{-} \eta) < 3.4 \times 10^{-7}$ 
(90 \% CL) \cite{expTauMuEtaBelle}, which yields 
\begin{eqnarray} 
 |\kappa_{32}|^2  \lsim   
0.3 \times 10^{-6} \times 
  \left(\frac{m_{A}^{}}{150 {\rm GeV}}\right)^4
  \left(\frac{60}{\tan \beta}\right)^{6},
 \label{lim}
\end{eqnarray} 
for $m_{A}^{} \sim m_{H}^{}$. The bound becomes relaxed for 
greater $m_A^{}$ and smaller $\tan\beta$ values. 

Next, we discuss theory predictions 
on the LF violating parameter $\kappa_{32}$ in the framework of the MSSM. 
Nonzero values of $\epsilon_{ij}$ then arise from the 
radiative correction due to the slepton mixing. 
They are calculated in the mass insertion method 
as
$\epsilon_{ij} \equiv (\epsilon_{1})_{i} \delta_{ij} +
 (\epsilon_{2})_{ij}$ \cite{babu,ellis,rossiNew,mu-e,rossi},
with   
\begin{align}
 (\epsilon_{1})_{i} =& 
 - \frac{\alpha'}{8\pi} \mu M_{1} 
  \left[
   2 I_{3} (M_{1}^{2}, m_{\tilde{l}_{Li}}^{2}, m_{\tilde{e}_{Ri}}^{2})
   +  I_{3} (M_{1}^{2},\mu^{2}, m_{\tilde{l}_{Li}}^{2})
   -2 I_{3} (M_{1}^{2},\mu^{2}, m_{\tilde{e}_{Ri}}^{2})
  \right] \nonumber \\
 &+
 \frac{\alpha_{2}}{8\pi} \mu M_{2}
 \left[
    I_{3} (M_{2}^{2},\mu^{2}, m_{\tilde{l}_{Li}}^{2})
 +2 I_{3} (M_{2}^{2},\mu^{2}, m_{\tilde{\nu}_{Li}}^{2})
 \right], \label{eq:eps1}\\
(\epsilon_{2})_{ij} =&
 -
 \frac{\alpha'}{8\pi} 
 \left(\Delta m_{\tilde{l}_{L}}^{2} \right)_{ij}
 \mu M_{1}
  \left[
  2 I_{4} (M_{1}^{2}, 
            m_{\tilde{l}_{Li}}^{2}, 
            m_{\tilde{e}_{Ri}}^{2},
            m_{\tilde{l}_{Lj}}^{2})
  +  I_{4} (M_{1}^{2},
            \mu^{2},
            m_{\tilde{l}_{Li}}^{2},
            m_{\tilde{l}_{Lj}}^{2})
  \right] \nonumber \\
 &+ \frac{\alpha_{2}}{8\pi} 
 \left(\Delta m_{\tilde{l}_{L}}^{2} \right)_{ij} \mu M_{2} 
 \left[
   I_{4} (M_{1}^{2},
            \mu^{2},
            m_{\tilde{l}_{Li}}^{2},
            m_{\tilde{l}_{Lj}}^{2})
 +
 2 I_{4} (M_{1}^{2},
            \mu^{2},
            m_{\tilde{\nu}_{Li}}^{2},
            m_{\tilde{\nu}_{Lj}}^{2})
 \right],\label{eq:eps2}
\end{align}
where $\alpha'$ and $\alpha_2$ are fine 
structure constants of $U(1)_{Y}$ and
$SU(2)_{L}$ symmetries, 
$M_{1}$ and $M_{2}$  
are the soft-SUSY-breaking masses for gauginos,  
$\mu$ is the SUSY-invariant Higgs
mixing parameter, and $m_{\tilde{l}_{Li}}^{2}$, $m_{\tilde{e}_{Ri}}^{2}$ 
and $m_{\tilde{\nu}_{Li}}^{2}$ are the left- and right-handed slepton 
and sneutrino masses of the $i$-th generation, respectively. 
The off-diagonal element of the slepton mass matrix is expressed by 
$\left(\Delta m_{\tilde{l}_{L}}^{2} \right)_{ij}$, $(i \neq j)$.\footnote{ 
We here assume the situation in which the origin of LFV is only the
mixing of the left-handed slepton.
The formulas which include the mixing of the right-handed slepton
are shown in Ref. \cite{rossi}.} 
The functions $I_{3}$ and $I_{4}$ are defined as
\begin{align}
I_{3} (x,y,z) \equiv&
 -
 \frac{xy \ln(x/y) + yz \ln(y/z) + zx \ln(z/x)}{(x-y)(y-z)(z-x)},\\
I_{4} (x,y,z,w) \equiv&
 -
 \frac{x \ln x}{(y-x)(z-x)(w-x)}
 -
 \frac{y \ln y}{(x-y)(z-y)(w-y)} \nonumber \\
 &-
 \frac{z \ln z}{(x-z)(y-z)(w-z)}
 -
 \frac{w \ln w}{(x-w)(y-w)(z-w)}.
\end{align}
Unlike the photon-mediation, 
the LF violating Yukawa coupling does not decouple 
for large values of the SUSY parameters.
It depends only on the ratio of the SUSY parameters.
For instance, by assuming  
$M_{1,2}=m_{\tilde{l}_{L\mu,\tau}}=m_{\tilde{\nu}_{L\mu,\tau}}
=m_{\tilde{e}_{R\mu,\tau}} = (\Delta m_{\tilde{l}_L^{}})_{32} \equiv 
m_{S} \neq \mu$, 
$(\epsilon_{1})_3$ and $(\epsilon_{2})_{32}$ in 
Eqs.~(\ref{eq:eps1}) and (\ref{eq:eps2}) are reduced to
\begin{align}
(\epsilon_{1})_{3} =&
 \frac{1}{8 \pi} R
 \left[
 -\alpha'
 +
 (\alpha'+3\alpha_{2}) \frac{R^{2} \ln R^{2} -R^{2} +1}{(R^{2} -1)^{2}}
 \right]
,\\
(\epsilon_{2})_{32} =&
 \frac{1}{8\pi}
 R
 \left[
 \frac{\alpha'}{3} + 
 \frac{\alpha' - 3 \alpha_{2} }{R^{2}-1}
 \left\{
 \frac{1}{2} - 
 \frac{R^{2} \ln R^2 -R^{2} +1}{(R^{2}-1)^{2}}
 \right\}
 \right],
\end{align}
where 
$R\equiv \mu/m_{S}$.
Therefore, magnitude of $|\kappa_{32}|^{2}$ becomes greater
as $R$ is larger. 

The photon-mediated LFV processes can be suppressed 
to be out of experimental reach when the typical 
SUSY breaking scale $m_{S}^{}$ is greater than ${\cal O}(1)$ TeV.
Let us consider the following choices. 
Case 1:
$\tan\beta=60$, $\mu=25$ TeV, 
$M_1 \sim M_2 \sim m_{\tilde{\ell}_{{L}_{\mu,\tau}^{}}}^{} \sim 
 m_{\tilde{\ell}_{{R}_{\mu,\tau}^{}}}^{} \sim
 m_{\tilde{\nu}_{L_{\mu,\tau}^{}}}^{} \sim  
 \left(\Delta m_{\tilde{l}_L^{}}^2 \right)_{32}^{} \sim 2$ 
TeV with the squark parameters 
$M_{Q}^{} \sim 10$ TeV and $M_{U,D}^{} \sim A_{t,b} \sim 8$ 
TeV. 
Case 2:
$\tan\beta=60$,  $\mu=10$ TeV,  
$m_{\tilde{\ell}_{{L}_{\mu,\tau}^{}}}^{} \sim
 m_{\tilde{\nu}_{L_{\mu,\tau}^{}}}^{} \sim 
\left(\Delta m_{\tilde{l}_L^{}}^2 \right)_{32}^{} \sim 1.2$ TeV, 
$m_{\tilde{\ell}_{{R}_{\mu,\tau}^{}}}^{} \sim 0.9$ TeV, 
$M_1 \sim 1$ TeV and $M_2 \sim 0.8$ TeV with the squark parameters 
$M_{Q}^{} \sim 5$ TeV and $M_{U,D}^{} \sim A_{t,b} \sim 3$ 
TeV.  
For Case 1 and Case 2, we obtain 
$|\kappa_{32}|^{2} \sim 8.4 \times 10^{-6}$ and $3.8 \times 10^{-6}$ 
with the gauge-mediated LF-violating processes being suppressed, 
respectively.
The branching fraction ${\rm Br}(h^0 \to \mu^\pm \tau^\mp)$ 
can be as large as $7 \times 10^{-4}$ for Case 1 with $m_A^{}=350$ GeV 
and $2 \times 10^{-4}$ for Case 2 with $m_A^{}=280$ GeV, respectively. 
We note that these extreme choices are not excluded 
by the condition of theory consistencies, such as 
color breaking, positiveness of eigenvalues of  
squark and slepton mass matrices. 

\begin{figure}[t]
\unitlength=1cm
\begin{picture}(10,7)
\hspace*{-0.5cm}
\includegraphics[width=12cm]{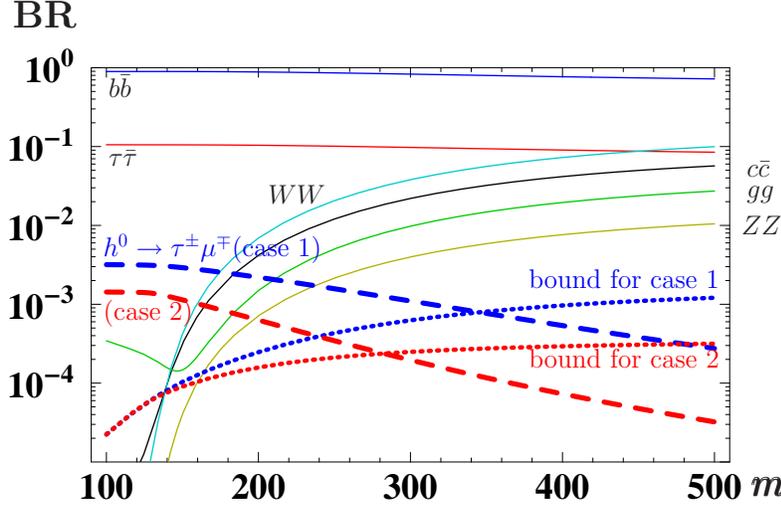}
\end{picture}
\vspace*{-5mm}
\caption{
The decay branching ratios of the lightest Higgs boson $h^0$ 
as a function of $m_A^{}$ at $\tan\beta=60$. 
The $m_h$ is set to be 123 GeV. 
The dashed curves represent the branching ratio 
for $h^0 \to \tau^\pm \mu^\mp$ in Case 1 and Case 2. 
The experimental upper constraint in Eq.~(\ref{lim}) is 
plotted for each case as a dotted curve. 
The branching ratios for the other decay modes are also shown 
for Case 1.
}
\label{Fig:Br-Mh}
\end{figure}

In Fig.~\ref{Fig:Br-Mh}, 
the decay branching ratio for the process $h^0 \to \tau^\pm \mu^\mp$ 
is shown as a function of $m_A^{}$ at $\tan\beta=60$. 
We take the other SUSY parameters so that the value of $m_h$ is 
123 GeV for each $m_A^{}$. 
The dashed curves represent ${\rm Br}(h^0 \to \tau^\pm \mu^\mp)$ 
in Case 1 and Case 2.
The experimental upper constraint in Eq.~(\ref{lim}) is 
also plotted as a dotted curve for each case. 
The branching ratio 
${\rm Br}(h^0 \to \tau^\pm \mu^\mp)$ can reach to 
$7 \times 10^{-4}$ and $2 \times 10^{-4}$ for Case 1 and Case 2, 
respectively.
In a wide region of $m_{A}^{}$, the branching ratio 
can be as large as $10^{-4}$ for both cases.

In the THDM, the parameters $\epsilon_{ij}$ in Eq.~\eqref{eq:L-Yukawa} 
can be taken freely within the experimental constraints and  
conditions from perturbative unitarity\cite{pu1,pu2} and 
vacuum stability\cite{vs}.  
The experimental bound on $|\kappa_{32}|$ can be weakened 
by considering the large value of $m_{A}^{}$ ($> 150$ GeV) 
and smaller $\tan\beta$ ($< 60$).
Therefore, much larger values of $|\kappa_{32}|$ are allowed 
in the THDM than those in the MSSM, especially for lower $\tan\beta$ 
values.

\section{Search for LF violating Higgs decays at a Linear Collider} 

\begin{figure}[t]
\unitlength=0.6cm
\begin{picture}(10,8)
\thicklines
 \put(1,1){\line(1,3){1}}
 \put(1,7){\line(1,-3){1}}
 \multiput(2,4)(0.5,0){3}{
  \qbezier(0,0)(0.125,0.25)(0.25,0)
  \qbezier(0.25,0)(0.375,-0.25)(0.5,0)}
 \put(3.5,4){\rotatebox{70}{
   \multiput(0,0)(0.5,0){3}{
   \qbezier(0,0)(0.125,0.25)(0.25,0)
   \qbezier(0.25,0)(0.375,-0.25)(0.5,0)}}}
  \put(4.15,5.4){\rotatebox{90}{\line(1,0){2}}}
  \put(4.15,5.4){\rotatebox{60}{\line(1,0){2}}}
 \multiput(3.5,4)(0.5,0){5}{\line(1,0){0.3}}
 \put(6,4){\line(1,3){1}}
 \put(6,4){\line(1,-3){1}}
  \put(6.42,5.3){\rotatebox{45}{\line(1,0){2}}}
  \put(6.42,5.3){\rotatebox{25}{\line(1,0){2}}}
 \put(2,4){\circle*{0.3}}
 \put(3.5,4){\circle*{0.3}}
 \put(6,4){\circle*{0.3}}
 \put(4.2,5.5){\circle*{0.3}}
 \put(6.47,5.35){\circle*{0.3}}
 \put(1.5,2.5){\vector(1,3){0}}
 \put(1.505,5.5){\vector(-1,3){0}}
 \put(6.5,2.5){\vector(-1,3){0}}
 \put(6.35,5){\vector(1,3){0}}
 \put(4.16,6.5){\vector(0,-1){0}}
 \put(4.82,6.5){\vector(2,3){0}}
 \put(6.85,6.5){\vector(1,3){0}}
 \put(7.4,6.25){\vector(1,1){0}}
 \put(7.4,5.75){\vector(-2,-1){0}}
 \put(0.3,0.3){$e^{-}$} \put(0.3,7.2){$e^{+}$}
 \put(2.5,3){$Z^{*}$} \put(3,5){$Z$}
 \put(4.5,3){$h^{0}$}
 \put(6.8,0.3){$\mu^{+}$} 
 \put(6.5,4.5){$\tau^{-}$} 
 \put(6.8,7.2){$\nu_{\tau}$}
 \put(8,6.5){\footnotesize $\tau$-decay}
 \put(3.5,7.4){\footnotesize $Z$-decay}
 \put(3.5,-0.5){(a)}
\end{picture}
\unitlength=0.6cm
\begin{picture}(10,8)
\thicklines
 \put(1,1){\line(1,3){1}}
 \put(1,7){\line(1,-3){1}}
 \multiput(2,4)(0.5,0){3}{
  \qbezier(0,0)(0.125,0.25)(0.25,0)
  \qbezier(0.25,0)(0.375,-0.25)(0.5,0)}
 \put(3.5,4){\rotatebox{70}{
   \multiput(0,0)(0.5,0){3}{
   \qbezier(0,0)(0.125,0.25)(0.25,0)
   \qbezier(0.25,0)(0.375,-0.25)(0.5,0)}}}
  \put(4.15,5.4){\rotatebox{90}{\line(1,0){2}}}
  \put(4.15,5.4){\rotatebox{60}{\line(1,0){2}}}
 \multiput(3.5,4)(0.5,0){5}{\line(1,0){0.3}}
 \put(6,4){\line(1,3){1}}
 \put(6,4){\line(1,-3){1}}
  \put(6.42,5.3){\rotatebox{45}{\line(1,0){2}}}
  \put(6.42,5.3){\rotatebox{25}{\line(1,0){2}}}
  \put(6.4,2.8){\rotatebox{-90}{\line(1,0){2}}}
  \put(6.4,2.8){\rotatebox{60}{\line(-1,0){2}}}
 \put(2,4){\circle*{0.3}}
 \put(3.5,4){\circle*{0.3}}
 \put(6,4){\circle*{0.3}}
 \put(4.15,5.4){\circle*{0.3}}
 \put(6.4,2.75){\circle*{0.3}}
 \put(6.47,5.4){\circle*{0.3}}
 \put(1.5,2.5){\vector(1,3){0}}
 \put(1.505,5.5){\vector(-1,3){0}}
 \put(6.15,3.5){\vector(-1,3){0}}
 \put(6.35,5){\vector(1,3){0}}
 \put(4.16,6.5){\vector(0,-1){0}}
 \put(4.82,6.5){\vector(2,3){0}}
 \put(6.85,6.5){\vector(1,3){0}}
 \put(7.4,6.25){\vector(1,1){0}}
 \put(7.4,5.75){\vector(-2,-1){0}}
 \put(6.655,2){\vector(-1,3){0}}
 \put(6.41,1.6){\vector(0,1){0}}
 \put(5.8,1.745){\vector(-2,-3){0}}
 \put(0.3,0.3){$e^{-}$} \put(0.3,7.2){$e^{+}$}
 \put(2.5,3){$Z^{*}$} 
 \put(3,4.5){$Z$} 
 \put(4.5,3){$h^{0}$}
 \put(6.8,7.2){$\nu_{\tau}$}
 \put(6.5,4.5){$\tau^{-}$}
 \put(6.5,3){$\tau^{+}$}
 \put(6.85,0.55){$\mu^{+}$}
 \put(6,0.35){$\bar{\nu}_{\mu}$} 
 \put(5,0.5){$\nu_{\tau}$}
 \put(8,6.5){\footnotesize $\tau$-decay}
 \put(3.5,7.4){\footnotesize $Z$-decay}
  \put(3.5,-0.5){(b)}
\end{picture}
\caption{The Feynman diagram of the signal event (a), and that of {\it the fake event} (b).}
\label{Fig:diagrams}
\end{figure}
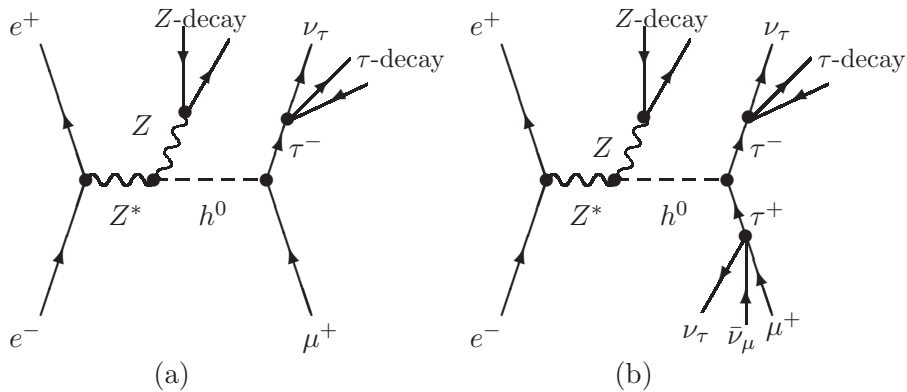

Let us consider the LF violating Higgs decay 
$h^{0} \to \tau^{\pm} \mu^{\mp}$ at a LC 
in the situation where the heavier Higgs bosons 
nearly
decouple from the gauge bosons; i.e.,
$\sin(\alpha-\beta) \simeq -1$. 
The lightest Higgs boson then approximately behaves as the SM one.  
The main production modes of the lightest Higgs boson at a LC
are the Higgsstrahlung $e^+e^- \to Z^\ast \to Z h^{0}$ and the $W$ fusion 
$e^+e^- \to (W^{+\ast} \bar{\nu}_{e})(W^{-\ast} \nu_{e}) 
\to h^{0} \nu_{e} \bar{\nu}_{e}$. 
For a light $h^0$ with the mass $m_h \sim 120$ GeV, 
the former production mechanism is dominant at low collision 
energies ($\sqrt{s} < 400$-$500$ GeV), while the latter dominates 
at higher energies. 
For our purpose, the Higgsstrahlung process 
is useful because of its simple kinematic structure. 
The signal process is then 
$e^+e^- \to Z^\ast \to Z h^{0} \to Z \tau^\pm \mu^\mp$.
We can detect the outgoing muon with high efficiency, 
and its momentum can be measured precisely by event-by-event. 
The momentum of the $Z$ boson can be reconstructed from those of 
its leptonic $\ell^+\ell^-$ 
($\ell^\pm = e^\pm$ and $\mu^\pm$) or hadronic ($jj$) decay products. 
Therefore, we can identify the signal event
without measuring $\tau$ momentum directly, 
as long as the beam spread rate for $\sqrt{s}$ 
is sufficiently low.

Depending on the $Z$ decay channel, the signal events are separated 
into two categories, 
$jj\tau^{\pm}\mu^{\mp}$ and $\ell^+\ell^-\tau^{\pm}\mu^{\mp}$.  
The energy resolution of the $Z$ boson 
from hadronic jets $jj$ is expected to
be $0.3\sqrt{E_{Z}}$ GeV   
and that from $\ell^+\ell^-$ is $0.1 \sqrt{E_{Z}}$ GeV\cite{lc}.
We assume that the detection efficiencies 
of the $Z$ boson and the muon are 100 \%,   
the rate of the beam energy spread is expected to be 0.1 \% level\cite{lc}, 
the muon momentum is measured with high precision 
and the mass of the lightest Higgs boson will have been determined 
in the 50 MeV level \cite{lc}. We also expect that the 
effect of the initial state radiation is small for the collider 
energies that we consider ($\sqrt{s} \sim 250$-$300$ GeV).
Taking into account all these numbers, we expect that the tau momentum 
can be determined indirectly within 3 GeV for 
$jj\tau^{\pm}\mu^{\mp}$ and 1 GeV for $\ell^+\ell^- \tau^{\pm}\mu^{\mp}$. 

Let us evaluate the number of the signal event. 
We assume that the energy $\sqrt{s}$ is tuned depending 
on the mass of the lightest Higgs boson: i.e.,
we take the optimal $\sqrt{s}$ to product the lightest Higgs boson
through the Higgsstrahlung process.
(It is approximately given by 
$\sqrt{s} \sim m_{Z} + \sqrt{2} m_{h}^{}$.)
The production cross section of $e^+e^- \to Z h^{0}$ is about 
$220$ fb for $m_{h}^{}=123$ GeV.
Then, we obtain $2.2 \times 10^5$ Higgs events 
if the integrated luminosity is 1 ab$^{-1}$.
When $|\kappa_{32}|^2 $ is 
$8.4 \times 10^{-6}$ (see Eq.~\eqref{lim}), 
about $118$ events of $jj \tau^\pm\mu^\mp$ and 
$11$ events of $\ell^+\ell^-\tau^{\pm}\mu^{\mp}$ 
can be produced. 
 
Next, we consider the background.  
For the signal with the Higgs boson mass of 120 GeV, 
the main background comes from $e^+e^- \to Z \tau^+\tau^- $. 
The number of the $Z \tau^{\pm}\mu^{\mp}$ event from 
$e^+e^- \to Z \tau^+\tau^-$ is estimated 
about $3.6 \times 10^{4}$\cite{comphep}.
Although the number of the background events is huge,
we can expect that a large part of them is effectively 
suppressed by using the following kinematic cuts:
(i) The muon from the Higgs boson should have high energies larger 
than $\sqrt{s}/4$, while those of the muon from the other parent 
are normally smaller.  Therefore, we impose 
the cut $E_\mu > \sqrt{s}/4$. 
(ii) The invariant mass $M_{\mu\tau}^{}$ distribution of the signal 
event (which is reconstructed from the 
information of the beam spread rate of $\sqrt{s}$ as well as 
the momenta of the outgoing muon and the $Z$ boson) 
should be located at the mass of the lightest Higgs boson, 
while that of the background is widely distributed.
By taking only events which satisfy 
$|M_{\mu\tau}^{} - m_h | < {\rm Max}[ 
\Gamma_h, \Delta {m_h}, \Delta M_{\mu\tau}^{} ]$,
the background events are expected to be considerably reduced, 
where $\Gamma_h$ ($\sim 40$ MeV for $m_h=120$ GeV) 
is the natural width of $h^0$, $\Delta m_h$ is the experimental 
uncertainty of $m_h$ ($\sim 50$ MeV), and 
$\Delta M_{\mu\tau}^{}$ is the uncertainty of 
the recoil invariant mass $M_{\mu\tau}^{}$. 
We here assume that $\Delta M_{\mu\tau}^{}$ is 1 GeV for the 
$Z \to \ell^+\ell^-$ channel and $3$ GeV for 
$Z \to jj$, respectively.

The irreducible background comes from the process shown in 
Fig.\ref{Fig:diagrams}-(b): 
the Higgs boson decays into a
tau pair, and one of the tau decays into a muon and missings ($e^{+}
e^{-} \rightarrow Z h^{0} \rightarrow Z \tau^{+}\tau^{-} \rightarrow
Z \tau^{\pm}\mu^{\mp}$+missings).
We can not distinguish the signal event $h^{0}\rightarrow \tau^{\pm}\mu^{\mp}$
with the event of Fig.\ref{Fig:diagrams}-(b) when the muon emitted 
from the tau lepton carries 
the similar momentum to that of the parent, 
because it leaves the same track on the detector as the signal event.
We refer this kind of the background as {\it the fake signal}. 
In the following, we estimate the number of the fake signal.
As the branching ratio for $h^{0} \rightarrow \tau^{+} \tau^{-}$
is about $0.1$, the initial number of the background event for  
$jj \tau^{\pm} \mu^{\mp}$+missings is calculated to be 
about 5200, 
and that for $\ell^+\ell^-\tau^{\pm}\mu^{\mp}$+missings is to be 500.
Since the signal includes the two-body decay of the Higgs boson  
$h^{0} \rightarrow \tau^{\pm} \mu^{\mp}$,
its muon energy distribution shows the mono-energetic spectrum.
On the other hand, that of the background, $h^{0} \rightarrow
\tau^{+}\tau^{-} \rightarrow \tau^{\pm} \mu^{\mp}+$missings, 
is the continuous spectrum.
The energy and angular distribution of the muon from the tau 
lepton in the lab frame is calculated as
\begin{align}
 \hspace*{-0.4cm}
\frac{{\rm d} n_{\mu}}{{\rm d}x {\rm d}\cos\theta_{h\mu}}
\! \simeq \!
 64 \gamma_{\tau}^{6} \gamma_{h}^{4} (1-\beta_{\tau})^{3}
 \left( 1 - \beta_{h} \cos \theta_{h \mu} \right)
 x^{2} \!
 \left\{
  3 - 
  8 \gamma_{\tau}^{2} \gamma_{h}^{2} 
    (1-\beta_{\tau})
    \left(
     1 - \beta_{h} \cos \theta_{h \mu} 
    \right) x
 \right\},
\label{eq:fake-fraction}
\end{align}
where $\gamma_{\tau} \equiv m_{h}/ (2 m_{\tau}) $ and $\beta_{\tau} \equiv
\sqrt{1-1/\gamma_{\tau}^{2}}$ are the boost factors from the
tau-rest frame to the Higgs-rest frame,
$\gamma_{h} \equiv E_{h}/m_{h}$ and $\beta_{h}\equiv
\sqrt{1-1/\gamma_{h}^{2}}$  are the boost factors from the
Higgs-rest frame to the lab frame,
$\theta_{h\mu}$ is the angle between momenta of the Higgs boson 
and the muon, 
and $x$ is defined as the ratio of the energy of the muon and that of
the parent tau lepton, $x\equiv E_{\mu}/E_{\tau}$.
Eq.~\eqref{eq:fake-fraction} 
can be derived from the differential cross section for
$\tau^{-}\rightarrow \mu^{-}\nu_{\tau}\bar{\nu}_{\mu}$ in the tau-rest
frame by making the boost twice.
In the boost from the Higgs-rest frame to the lab frame,
we take the approximation in which the muon is emitted to the
forward direction of the tau lepton.
The number of events of the fake signal can be evaluated as
\begin{align}
N_{\text{fake}} =  N_{Z\mu\tau}^\text{initial} 
 \times
 \int_{\theta_{h\mu} = 0}^{\theta_{h\mu} =\pi}
 {\rm d}\cos\theta_{h\mu}
 \int^{x_{\text{max}}}_{x_{\text{max}} - \delta x}
 {\rm d}x
 \frac{{\rm d} n_{\mu}}{{\rm d}x {\rm d}\cos\theta_{h\mu}},
 \label{eq:number-of-fake}
\end{align}  
where $N_{Z\mu\tau}^\text{initial}$ is 
the initial number of the background event for 
$Z\mu\tau$ with $Z \to jj$ or $Z \to \ell^+\ell^-$, 
$x_{\text{max}}$ is the maximal value of $x$ which is given by 
\begin{align}
x_{\text{max}} \equiv 1/\left\{ 4 \gamma_{\tau}^{2} \gamma_{h}^{2} (1-\beta_{\tau})
 (1-\beta_{h} \cos\theta_{h\mu}) \right\},
\end{align}
and parameter $\delta x$ depends on the uncertainty of the
tau momentum, $\delta(E_{\tau})$;
\begin{align}
\delta x \equiv \delta \left( \frac{E_{\mu}}{E_{\tau}} \right)
 \simeq x_{\text{max}}\frac{\delta (E_{\tau})}{E_{\tau}}.
\end{align}  
We find that the number of the fake signal strongly depends on the precision of the
tau momentum determination.
We expect that it is attained with the similar
precision to that of the Higgs boson mass 
reconstructed by the recoil momentum.
We here take the uncertainty of the tau momentum as 3 GeV
for $jj\tau^{\pm}\mu^{\mp}$ and as 1 GeV for 
$\ell^+\ell^- \tau^{\pm}\mu^{\mp}$. 

Finally, we estimate the statistical significance ($S/\sqrt{B}$) 
for each channel. The number of the fake events is evaluated by
Eq.~\eqref{eq:number-of-fake}, which is 
460 for $jj\tau^{\pm}\mu^{\mp}$ and 15 for 
$\ell^+\ell^- \tau^{\pm} \mu^{\mp}$.
Therefore, in Case 1 where $|\kappa_{32}|^2 $ is $8.4 \times 10^{-6}$ 
with $m_h^{}=123$ GeV, the significance can become 5.5 and 3.0 
for $jj \tau^\pm\mu^\mp$ and $\ell^+\ell^-\tau^{\pm}\mu^{\mp}$ at 
$m_A^{}=350$ GeV, 
respectively, taking into account the constraint from the
$\tau^- \to \mu^- \eta$ result given in Eq.~(\ref{lim}).
The combined significance can reach to 6.3. 
In Case 2 where $|\kappa_{32}|^2 $ is $3.8 \times 10^{-6}$ 
with $m_h^{}=123$ GeV,  
the number of the signal becomes smaller, and the combined 
significance amounts to be as large as 2.0 at $m_A^{}=280$ GeV.

\begin{figure}[t]
\unitlength=1cm
\begin{picture}(10,7)
\hspace*{-2cm}
\includegraphics[width=12cm]{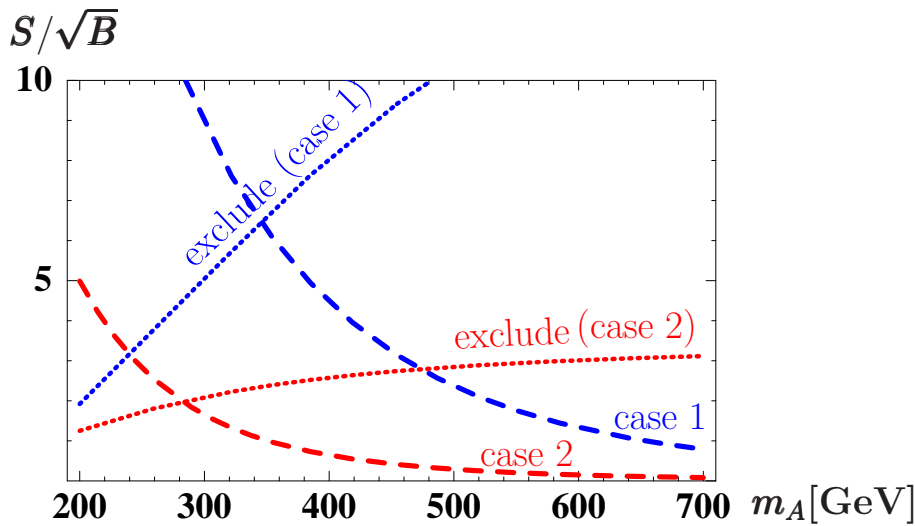}
\end{picture}
\vspace*{-5mm}
\caption{
The statistical significance $S/\sqrt{B}$ is plotted 
(dashed curve) as a function of $m_A^{}$ at $\tan\beta=60$ 
in Case 1 where $|\kappa_{32}|^2 \simeq 8.4 \times 10^{-6}$ 
and Case 2 where $|\kappa_{32}|^2 \simeq 3.8 \times 10^{-6}$ .  
The mass of the lightest Higgs boson $h^0$ is set to be 
$123$ GeV. The upper bound from the current data 
for $\tau^- \to \mu^- \eta$ is also shown as a dotted curve for 
each case. }
\label{Fig:Detectable-kSq}
\end{figure}

\section{Summary and discussions}
We have discussed detecting the lepton flavor violating 
decay mode of the Higgs boson $h^0 \to \tau^\pm \mu^\mp$ at a LC. 
The effective coupling of $h^0 \tau^\pm \mu^\mp$ is induced 
at one loop in the MSSM due to the slepton mixing. 
We have studied the situation where the typical scale of 
supersymmetric parameters is as large as TeV scale. 
The magnitude of the effective $h^0 \tau^\pm \mu^\mp$ coupling 
can then be substantially large. 
Consequently, the number of the signal event via 
$e^+e^- \to Z h^0 \to Z \tau^\pm \mu^\mp$ can be large 
enough to be detected after the background is suppressed 
by kinematic cuts. The signal can be marginally visible 
in the MSSM when the effective $h^0 \tau^\pm \mu^\mp$ coupling 
becomes enhanced due to the large ratio of $\mu$ and $m_{S}^{}$, 
where $m_{S}^{}$ is the typical scale of the soft-breaking mass.

When $m_{S}^{}$ is greater than the TeV scale, the LF violating 
processes associated with gauge bosons such as 
$\tau^- \to \mu^- \gamma$, $\tau^- \to e^- \gamma$ and 
$\mu^- \to e^- \gamma$ are suppressed. In addition, the LF violating 
processes including the Higgs mediation such as 
$\tau^- \to \mu^- \eta$, $\tau^- \to \mu^- \mu^+ \mu^-$ and 
$\mu^- N \to e^- N$ as well as the flavor changing processes 
such as $b \to s \gamma$ are suppressed when $m_A^{}$ is greater 
than about 300 GeV\cite{bsgamma}. 
On the other hand, the branching ratio for $h^0 \to \tau^\pm \mu^\mp$ 
does not decouple for large $m_{S}^{}$ as long as the ratio 
$\mu/m_{S}^{}$ is not small.  Therefore, in such a case, the decay 
$h^0 \to \tau^\pm \mu^\mp$ at a LC can be a complementary process 
to test the Higgs mediated LF violating coupling. 

We comment on the case of the general framework of the THDM. 
Unlike the MSSM, the mixing angle $\alpha$ is independent 
of $\tan\beta$ and $m_A^{}$. 
For larger values of $m_A$, the bound from $\tau^- \to \mu^- \eta$ 
can be relaxed by the factor of $m_A^4$ (cf. Eq.~(\ref{lim})), whereas 
the branching ratio of the $h^0 \to \tau^\pm \mu^\mp$ can remain 
be larger than $10^{-3}$ within the available experimental and 
theoretical constraints. Therefore, the number of the signal 
in the THDM can be by a few order of magnitude 
larger than the possible value in the MSSM. 

In our analysis, we have used the bound on the LF violating coupling 
$|\kappa_{32}^{}|$ from the current data of $\tau^- \to \mu^- \eta$.
In near future, if the bound becomes strong by a few factor, 
the number of the signal becomes reduced by the same factor.  
We have assumed some important numbers which 
are associated with the machine property for the collider and 
the detector of a LC experiment. 
The estimation of the number of the signal events and the 
reduction of the background events largely 
depends on the detection efficiencies of $Z$ and $\mu$, 
the resolution of the momenta for them, the rate of beam energy 
spread of the $e^+e^-$ collision and the initial state radiation. 
Our assumption for these numbers might be rather optimistic.
On the other hand, 
the significance can be improved when direct detection of the 
tau lepton is taken into account. 
In any case, a more realistic simulation analysis is 
necessary to determine feasibility of the signal.  

\vspace{1cm}
\noindent
{\large \it Acknowledgments}

The authors would like to thank  
Chikara Fukunaga, 
Masafumi Koike,
Yoshitaka Kuno,
Yasuhiro Okada, 
and 
Minoru Tanaka 
for valueable discussions. 
We also would like to thank 
Michael Peskin and Isamu Watanabe 
for useful suggestions and comments 
at the YITP Meeting on Phenomenology 
in June 2004.


\end{document}